\documentclass{ws-rv961x669}
\usepackage[square]{ws-rv-van}     
\usepackage{ws-rv-thm}     
\usepackage{subfigure}     
\makeindex

\usepackage{calrsfs}
\DeclareMathAlphabet{\pazocal}{OMS}{zplm}{m}{n}
\usepackage{cite} 

\allowdisplaybreaks[4]

\newcommand{\GeV}{\rm GeV}

 \newcommand{\Atil}{\tilde{A}}
 \newcommand{\Ctil}{\tilde{C}}
\begin{document}

\chapter[Deep-Inelastic Scattering: What do we know ?]{
Deep-Inelastic Scattering: What do we know ?}

\author{{Johannes Bl\"umlein}}

\address{Deutsches Elektronen-Synchrotron DESY, \\ Platanenallee 6, 15738 
Zeuthen, Germany\\
Johannes.Bluemlein@desy.de\footnote{DESY 23--066,~~DO--TH 23/06.}
}

\begin{abstract}
A survey is given on the current status of the theoretical description
of unpolarized and polarized deep--inelastic scattering processes in
Quantum Chromodynamics at large virtualities.
\end{abstract}

\begin{center}
{\sf Dedicated to the Memory of Harald Fritzsch.}
\end{center}


\body


\section{Introduction}\label{sec:1}

\vspace*{1mm}
\noindent
About 50 years ago Quantum Chromodynamics (QCD), the theory of strong
interactions, was found \cite{Nambu:1966,Fritzsch:1972jv,Gross:1973id,
Politzer:1973fx,Fritzsch:1973pi}, and Harald Fritzsch played a major role
in this. After the proof of renormalizibility \cite{Veltman:1963,'tHooft:1971fh}
of the Yang-Mills theories \cite{Yang:1954ek} and the proof of the anomaly freedom 
\cite{ANOMALYF} of the $SU_{2L} \times U_{1Y} \times SU_{3c}$ Standard Model, systematic 
perturbative calculations became possible to make predictions for experimental 
precision measurements of several hard scattering processes under certain kinematic 
conditions. The first calculation
concerned the running of the strong coupling constant \cite{Gross:1973id,
Politzer:1973fx} proving asymptotic freedom, a {\sf conditio sine qua non} for higher
order calculations in perturbative QCD. One of the
key processes is deeply inelastic scattering (DIS) of leptons off nucleons, both in the 
unpolarized and polarized case. These processes allow to determine the various
quark flavor and gluon densities 
\cite{PART,Jimenez-Delgado:2014twa,Alekhin:2017kpj,Blumlein:2006be}, reveal higher twist 
contributions
in the region of low virtualities $Q^2$ and/or large values of the Bjorken variable 
$x$,\cite{HT,Alekhin:2012ig}, with 
$x = Q^2/(2 p.q)$, $Q^2 = - q^2$, $q$ the 4-momentum transfer from the lepton 
to the nucleon, and $p$ the nucleon momentum. 

Deep--inelastic scattering has been the method
to observe scaling \cite{SCALING}, predicted in Ref.~\cite{Bjorken:1968dy} and leading to the parton 
model \cite{FEYNMAN}. It allows to measure the strong coupling constant $\alpha_s(M_Z^2)$ at high 
precision \cite{alphas} from the scaling violations of the deep--inelastic structure functions 
\cite{SCALVIOL}. Furthermore, their heavy flavor contributions allow a precision measurement of the
charm quark mass $m_c$ \cite{Alekhin:2012vu}.

In this survey we will sketch the way the QCD corrections to deep--inelastic scattering took
from the  beginning of the light-cone picture \cite{LCE} and (naive) parton model \cite{FEYNMAN}
until today, concerning the running of the coupling, the anomalous dimensions, the different massless
and massive Wilson coefficients in both the unpolarized and polarized case.\footnote{For earlier reviews 
see Refs.~\cite{Politzer:1974fr,REV,Blumlein:2012bf}.} The 
corrections to the QCD $\beta$--function, being a zero--scale 
quantity, are simpler to calculate than the anomalous dimensions for general values of the Mellin variable 
$N$, which are somewhat simpler than the massless Wilson coefficients, followed in complexity by analytic 
results on the massive Wilson coefficients.

For the unpolarized anomalous dimensions the first order corrections were known in 1974, the second 
order 
corrections in 1980 (with a final correction in 1991), and the third order results in 2004.
The main massless Wilson coefficients were known in correct form 1980 at one loop, in 1992 at two loops 
and 2005 at three loops. In the massive case the one loop results were known in 1976, the two-loop results
emerged between 1992 and 1996, and the asymptotic three loop results between 2010 and today, in the single 
and two--mass cases. The first fixed Mellin moments of these quantities were available earlier, if they 
had been calculated, and a series of lower moments has been calculated  at four--loop order since 2016. 
The QCD $\beta$ function
is known to five--loop order since 2016/17. This time--line shows the challenge, which in part had to 
involve new mathematical developments and certainly great efforts in computer algebra.

We will describe this development for the anomalous dimensions, the Wilson coefficients and the 
renormalization-group quantities in the following, which form the asset needed to describe the scaling 
violations of the deep--inelastic structure functions. Furthermore, we will add some remarks on the 
Drell--Yan process, since these data are needed to fix the light sea--quark distributions in QCD 
analyzes. We will also comment on technical and mathematical challenges connected to these analytic 
calculations, which became indispensable since the time of about 1998 and were essential to achieve the 
present status.
\section{Scaling violations of DIS structure functions}
\label{sec:2}

\vspace*{1mm}
\noindent
The measurement of the fundamental parameters of the Standard Model, such as $\alpha_s(M_Z^2)$ or the 
heavy quark masses $m_Q$, require clear conditions. One has to choose a kinematic region for which 
definite theoretical predictions can be made. Here one request is that non--perturbative and perturbative
effects can be clearly separated and the perturbative corrections can be carried out safely. In 
deep--inelastic scattering one is therefore advised to consider the kinematic region of the dominance 
of twist\cite{Gross:1971wn} $\tau = 2$ operators, which means that the virtuality $Q^2$ of the process
must be large to approach the Bjorken limit \cite{Bjorken:1968dy}. The effect of the higher twist 
contributions \cite{HT,Alekhin:2012ig}\footnote{See also Section~16 of Ref.~\cite{Blumlein:2012bf}.} is 
then suppressed.
One might choose to include only data with $Q^2 > 25$~GeV$^2$ and $W^2 = Q^2 (1-x)/x > 12.5$~GeV$^2$  
into the analysis, despite the fact of 
a high statistics measured below these scales, to avoid biases of not completely known power- and other 
corrections. It is known, cf. Ref.~\cite{Alekhin:2012ig}, that one may obtain {\it different} Standard 
Model 
parameters if these cuts are weakened pointing to corrections not being controlled.

Moreover, 
one may use the asymptotic heavy quark Wilson coefficients in this region \cite{Buza:1995ie}, 
which can be calculated analytically, cf.~Section~\ref{sec:6}. Under these conditions one obtains the 
single particle factorization \cite{FACT}  between the process-dependent Wilson coefficients and the 
single parton 
distribution functions, which is otherwise not the case. In going systematically to higher and higher 
orders there are also no limitations in approaching the small $x$ region, since the calculation within
QCD is complete. Small $x$ approaches have to rely on factorization between the perturbative and 
non--perturbative contributions \cite{FACT} as well and refer to  the twist expansion for consistent
renormalization. As has been shown in \cite{Blumlein:1997em,Blumlein:1999ev}, several sub--leading 
small $x$ series have to be known to obtain results which are phenomenologically stable.
In the following we will concentrate on the theoretical calculations under the above 
conditions.

Let us consider the dynamics in Mellin $N$ space. The Mellin transform of a function $f(x)$ is given by
\begin{eqnarray}
f(N) = \int_0^1 dx x^{N-1}~\hat{f}(x).
\end{eqnarray}
In the twist--2 approximation the deep--inelastic structure functions $F_i$ obey the representation
\begin{eqnarray}
F_i(N,Q^2) = \sum_{l = q,g} C_{i,l}(N,a_s,Q^2/\mu^2) \cdot f_l(N, a_s), 
\label{eq:SFi}
\end{eqnarray}
where $C_{i,l}(N,Q^2/\mu^2)$ denote the renormalized Wilson coefficients, $f_l$ the renormalized
parton distribution functions, and $a_s = \alpha_s/(4\pi)$. Introducing the  operator\cite{Symanzik:1970rt}
\begin{eqnarray}
\label{eq:ren1}
{\mathcal D}(\mu^2) := \mu^2 \frac{\partial}{\partial \mu^2}
                 + \beta(a_s(\mu^2)) \frac{\partial}{\partial a_s(\mu^2)}
                 - \gamma_m(a_s(\mu^2)) m(\mu^2) \frac{\partial}{\partial m(\mu^2)},
\end{eqnarray}
with
\begin{eqnarray}
\label{eq:ren1a}
\beta(a_s(\mu^2))    &=&   \mu^2 \frac{\partial a_s(\mu^2)}{\partial \mu^2},~~~
\gamma_m(a_s(\mu^2)) = - \frac{\mu^2}{m(\mu^2)} \frac{\partial m(\mu^2)}{\partial \mu^2}~,
\end{eqnarray}
one obtains the following renormalization group equations (RGEs) from (\ref{eq:SFi}) 
\begin{eqnarray}
\label{eq:ren3}
\sum_j \left[{\cal D}(\mu^2) \delta_{ij} + \gamma_{ij}^{\rm S,NS}
- n_\psi \gamma_\psi - n_A \gamma_A
\right] f_j(N,\mu^2) &=& 0, \\
\label{eq:ren4}
\sum_j \left[{\cal D}(\mu^2) \delta_{ij} + \gamma_{J_1} + \gamma_{J_2} - \gamma_{ij}^{\rm S,NS}\right]
C_i\left(N,\frac{Q^2}{\mu^2}\right)
&=& 0~.
\end{eqnarray}
Here $\gamma_\psi$, $\gamma_A$ and $\gamma_{J_{1,2}}$ denote the anomalous dimension of external 
quarks, gluons, and the currents, which can be non-zero if the currents are not conserved.
$\gamma_{ij}^{\rm S,NS}$ denote the anomalous dimensions of the local operators 
(\ref{eq:op1}, \ref{eq:op2}). The scale dependence is due to $a_s(\mu^2)$, 
\begin{eqnarray}
\label{eq:run}
\mu^2 \frac{d a_s(\mu^2)}{d \mu^2} = - \sum_{k=0}^\infty \beta_k a_s^{k+2}(\mu^2), 
\end{eqnarray}
where one finally considers $\mu^2 = Q^2$.
\section{Zero scale quantities}
\label{sec:3}

\vspace*{1mm}
\noindent
The renormalization group equations 
for the massless and massive 
operator matrix elements and the 
Wilson coefficients also describe the scale dependence of the strong coupling constant and of the 
heavy quark masses. These are zero scale quantities and they have to be calculated to the respective
order in perturbation theory. The running of the coupling constant $\alpha_s(\mu^2)$ is described by
(\ref{eq:run}). Similar equations hold for the other quantities. In the $\overline{\rm MS}$ scheme one 
may
compute the different $Z$-factors renormalizing QCD in the case that there are no composite operators.
 
The one--loop result for the QCD $\beta$--function has been calculated in 
Refs.~\cite{Gross:1973id,Politzer:1973fx}, the two--loop corrections in Refs.~\cite{BETA2},
the three--loop contributions in Refs.~\cite{BETA3}, the four--loop corrections in 
Refs.~\cite{BETA4}, and most recently the five--loop corrections in Refs.~\cite{BETA5}.
The effect of asymptotic freedom observed in \cite{Gross:1973id,Politzer:1973fx} is refined by the higher 
order corrections for the number of active quark flavors $N_F \leq 6$ of nature.
The other $Z$--factors to renormalize QCD were calculated in higher orders in
Refs.~\cite{Egorian:1978zx,Vermaseren:1997fq,Chetyrkin:2004mf,Chetyrkin:2017bjc,Luthe:2016xec} and are now 
also available at five--loop order. The running of the heavy quark masses and the wave function 
renormalization have  been calculated in 
Refs.~\cite{MASSREN}. Herewith all necessary $Z$-factors occurring in processes without local operators 
are known.
\section{Anomalous dimensions and splitting functions}
\label{sec:4}

\vspace*{1mm}
\noindent
The QCD evolution of the twist--2 parton densities is ruled by the 
anomalous dimensions in $N$ space, $\gamma_{ij}(N)$, or the splitting 
functions in $x$ space, where the latter are the inverse Mellin 
transform of the former. The evolution equations derive from the RGE~(\ref{eq:ren3}) in non--singlet 
and singlet cases,
\begin{eqnarray}
\frac{d f_i^{\rm NS}(N,\mu^2)}{d \ln(\mu^2)} ~&=&~ - \gamma_{qq}^{\rm NS}(N, a_s) f_i^{\rm 
NS}(N,\mu^2),
\\
\frac{d}{d \ln(\mu^2)}  
\Biggl[
\begin{array}{c}
\Sigma(N,\mu^2) \\ G(N,\mu^2)
\end{array} \Biggr] 
~&=&~ - \Biggl[
\begin{array}{cc}
\gamma_{qq}(N, a_s) &
\gamma_{qg}(N, a_s) \\
\gamma_{gq}(N, a_s) &
\gamma_{gg}(N, a_s)
\end{array}
\Biggr]
\Biggl[ 
\begin{array}{c}
\Sigma(N,\mu^2) \\ G(N,\mu^2)
\end{array} 
\Biggr],
\end{eqnarray}
with $i = 1,2,3$ and $a_s = a_s(\mu^2)$.
The twist-2 parton densities are given as forward matrix elements $\langle i|O_k| j \rangle$ of the 
composite operators
\begin{eqnarray}
\label{eq:op1}
O^{\sf NS}_{q;r;\mu_1, \ldots, \mu_n}(0) &=& i^{N-1}{\bf S} \left[\bar{\psi} \gamma_{\mu_1}
D_{\mu_2} \ldots D_{\mu_N} \frac{\lambda_r}{2}\psi\right],
\\
\label{eq:op2}
O^{\sf S}_{q;r;\mu_1, \ldots, \mu_n}(0) &=& i^{N-1}{\bf S} \left[\bar{\psi}  \gamma_{\mu_1}
D_{\mu_2} \ldots D_{\mu_N} \psi\right],
\\
\label{eq:op3}
O^{\sf S}_{g;r;\mu_1, \ldots, \mu_n}(0) &=& 2 i^{N-2}
{\bf S} {\bf Sp} \left[ F^a_{\beta \gamma} D_{\mu_2} \ldots D_{\mu_{N-1}}
F^{\alpha,a}_{\mu_N}\right],
\end{eqnarray}
in the unpolarized case (with similar expressions in the polarized case).
Here the indices $q$ and $g$ refer to quark and gluon field operators, respectively, and $\lambda_r$ 
denotes
the Gell-Mann matrix of the corresponding light flavor representation; $\psi$
is the quark field, $D_{\mu}$ the covariant derivative, $F^a_{\alpha \beta}$ the Yang--Mills
field strength tensor,
{\bf S} the symmetry operator for all Lorentz indices and {\bf Sp} the color-trace, where the index $a$ 
is the color index in the adjoint representation. 
One calculates both the fixed Mellin moments using certain techniques
as well as the complete functions for general values of $N$. Here 
either the even or odd values of $N$ contribute, depending on the respective 
amplitude crossing relations, cf.~\cite{Politzer:1974fr,Blumlein:1996vs}.
\subsection{Fixed Moments}
\label{sec:41}

\vspace*{1mm}
\noindent
The first information one obtains on the anomalous dimensions is given by their fixed moments.
One may calculate them by differentiating the forward Compton amplitude by the proton momentum $p$.
In this way one works without reference, however, equivalent to the local twist two operators. 
The method has the advantage 
that also the moments of the respective massless Wilson coefficients can be obtained in this way.
In the massless case the {\tt Mincer} algorithm has been used \cite{MINCER} to three--loop order. In 
the massive case 
one uses the package {\tt MATAD} \cite{Steinhauser:2000ry}.
At two--loop order this has been done in Ref.~\cite{Larin:1991fv}. The method has later been expanded 
to three--loop order in Refs.~\cite{MOM3}, reaching an intermediate technical limit calculating the 16th
moment in the flavor non--singlet case in 2003. In the case of massive operator matrix elements 
(OMEs) moments between $N=10$ and $N=16$ were calculated in Ref.~\cite{Bierenbaum:2009mv}
at three--loop order. In the two--mass case moments were calculated in 
Refs.~\cite{Ablinger:2017err}.
The method implies an exponential rise of terms to 
be calculated and therefore terminates at a given order, depending on the complexity of the given 
problem. 

More recently, also a series of lower moments at four-- and five--loop order  have been calculated
by basically the same method in Refs.~\cite{MOM45,Davies:2016jie}, using {\tt Forcer} 
\cite{Ruijl:2017cxj} now having 
reached $N = 20$ in the four--loop case. This provides the most far reaching information at 
the moment. For simpler structures the number of moments obtained allow the reconstruction of the 
general $N$ results under certain assumptions \cite{MOM45}, in particular also that only 
harmonic sums \cite{HSUM} contribute. 

As known from the light--cone expansion \cite{LCE}, the Mellin moments are the genuine quantities in 
describing the scaling violations of DIS structure functions. In any approach to the calculation 
of DIS anomalous dimensions and Wilson coefficients one may transform the integration-by-parts 
identities (IBP) \cite{IBP} into difference equations for the master integrals, and related to that, 
the
amplitude. By using the method of arbitrarily high moments \cite{Blumlein:2017dxp} one may calculate 
very effectively high numbers of moments. Even in the massive case we have  generated 15.000 
moments at three--loop order recently.

The method of Ref.~\cite{Blumlein:2017dxp} allows then to use the method of guessing \cite{GUESS}
to find the associated recurrence, given one has had a sufficient number of moments. Here no special 
assumptions on the mathematical structure of the results are made unlike 
in the approach used e.g. in Ref.~\cite{MOM45}. The obtained
recurrences are then inspected using difference ring theory algorithms as implemented in the package
{\tt Sigma} \cite{SIG}. In the case of first order factorizing problems the general $N$ solution
can be calculated. In all other cases the first order factors can be separated off. This method
could be applied to all anomalous dimensions and massless Wilson coefficients to three loop order,
which are first--order factorizable problems. This also applies to various massive Wilson coefficients
at three--loop order, as will be discussed below. 

The calculation of the Mellin moments by using the differentiation method  is very different form other 
approaches. Therefore these results provide firm checks on the results for the case of general $N$ 
recurrences, without making special assumptions. 
\subsection{Results at General \boldmath $N$}
\label{sec:42}

\vspace*{1mm}
\noindent
The leading order unpolarized anomalous dimensions were calculated in Refs.~\cite{SCALVIOL}
and in the polarized case in \cite{PO1}. A partonic approach has been used in Ref.~\cite{Parisi:1976qj}, 
which is related to Refs.~\cite{SCALVIOL,PO1} by a Mellin transform.\footnote{For further one--loop results see
Ref.~\cite{Blumlein:2012bf}, Section~7.} The next-to-leading order anomalous dimensions and splitting 
functions were computed in Refs.~\cite{NLOu,Moch:1999eb} resp. in Refs.~\cite{NLOp}. Finally, the 
next-to-next-to-leading order ones in Refs.~\cite{Moch:2004pa,Vermaseren:2005qc,Blumlein:2021enk,
Blumlein:2022gpp,Ablinger:2014lka,Ablinger:2017tan,HADR,Duhr:2020seh,Gehrmann:2023ksf} and  
Refs.~\cite{Moch:2014sna,Behring:2019tus,
Blumlein:2021ryt,Blumlein:2022gpp} in the unpolarized and polarized cases. Simpler color factors
at four--loops are available at general values of $N$ \cite{Davies:2016jie,GEHR23}. 
Here different techniques have been used, such as the forward Compton amplitude \cite{Moch:2004pa,
Vermaseren:2005qc,Moch:2014sna,Blumlein:2022gpp}, massive on--shell OMEs 
\cite{Ablinger:2014lka,Ablinger:2017tan,Behring:2019tus} massless off--shell 
OMEs \cite{Blumlein:2021enk,Blumlein:2021ryt,Gehrmann:2023ksf}, and different 
hard scattering cross sections with on--shell amplitudes \cite{HADR,Duhr:2020seh}.
All contributions can be expressed in terms of harmonic sums or in 
$x$--space by the corresponding Mellin inversion in terms of harmonic 
polylogarithms \cite{Remiddi:1999ew}.
\begin{figure}
\centering
    \includegraphics[width=0.40\textwidth]{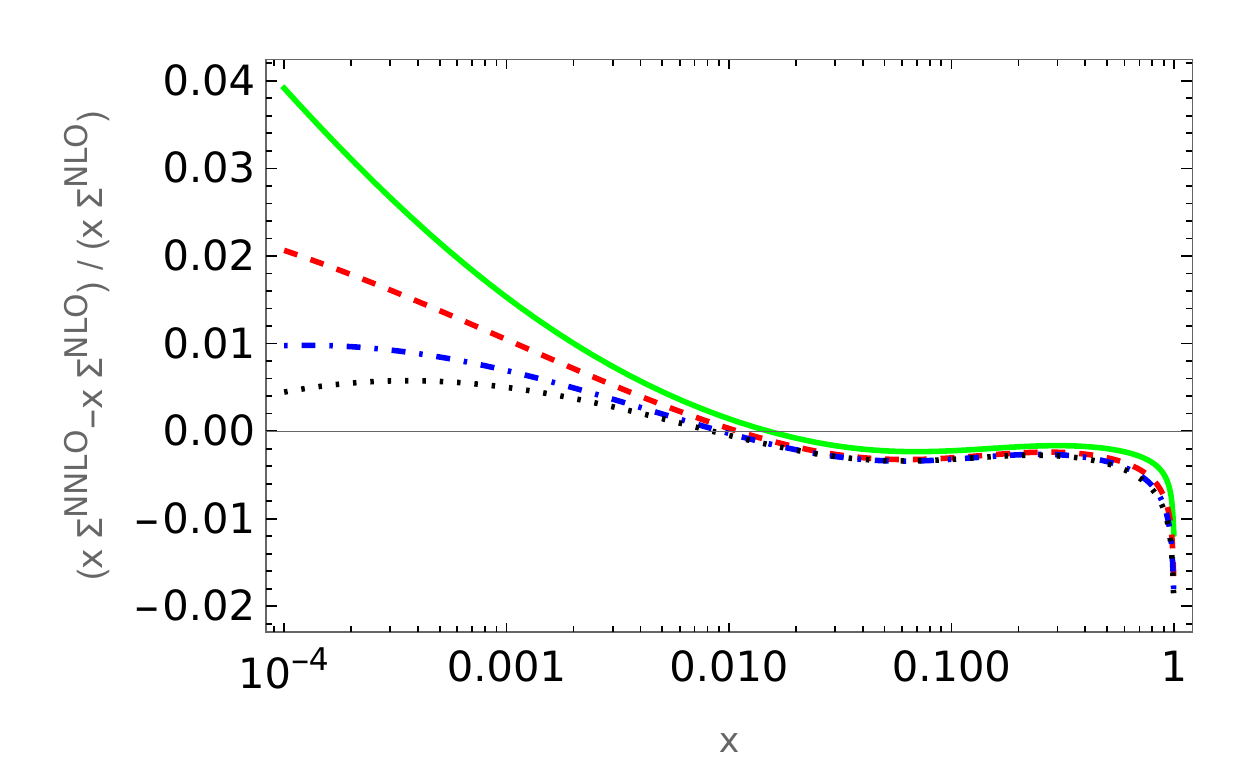} 
    \includegraphics[width=0.40\textwidth]{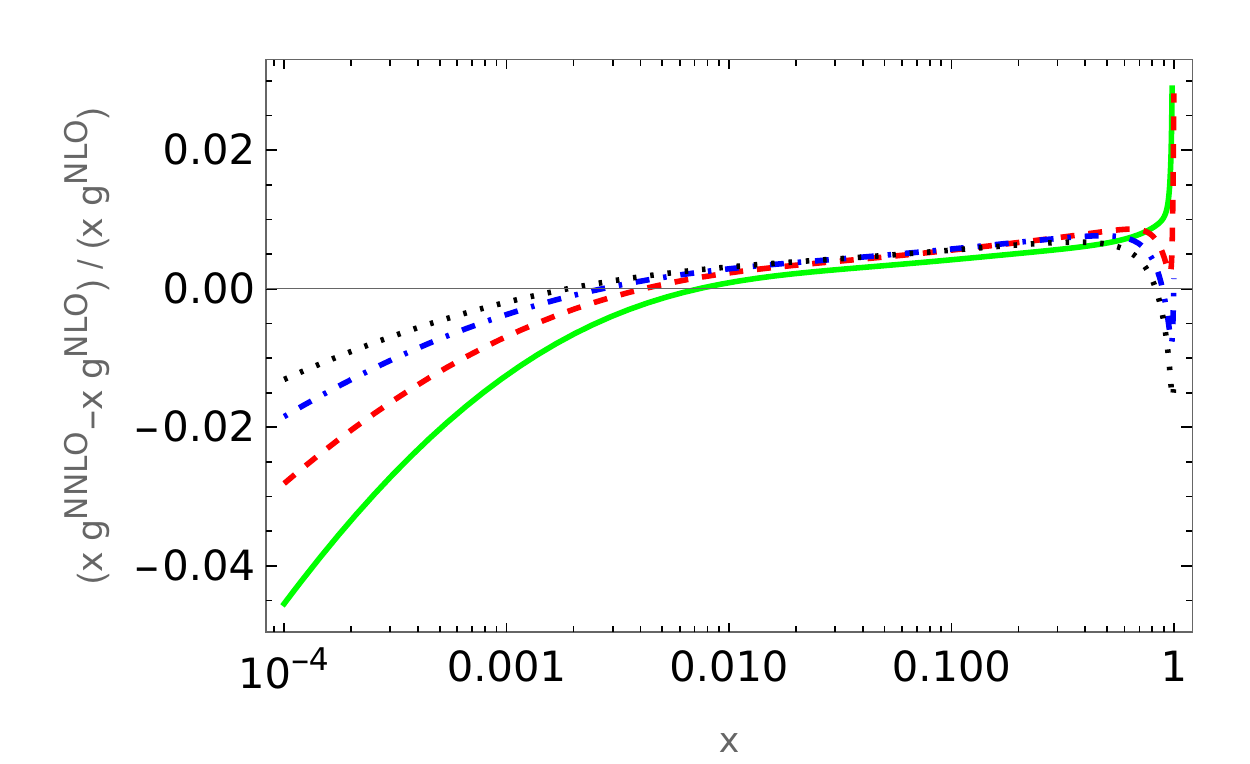} 
    \includegraphics[width=0.40\textwidth]{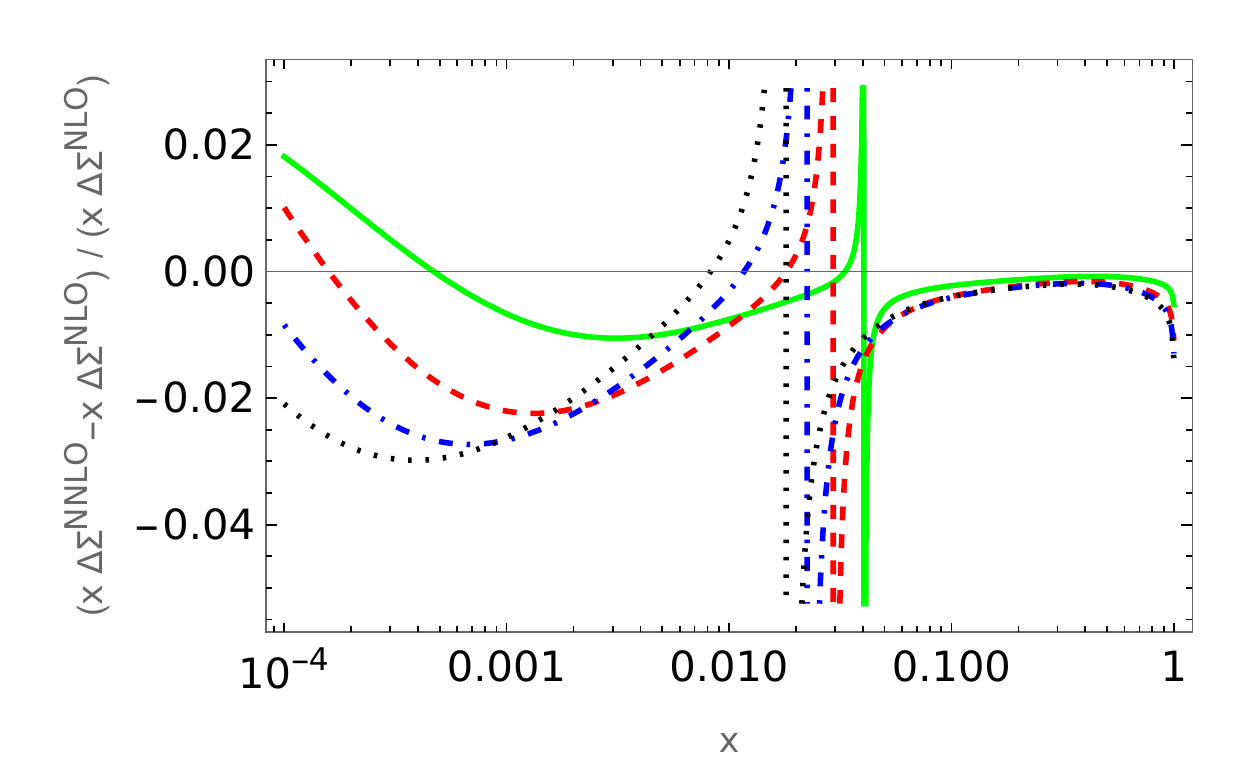} 
    \includegraphics[width=0.40\textwidth]{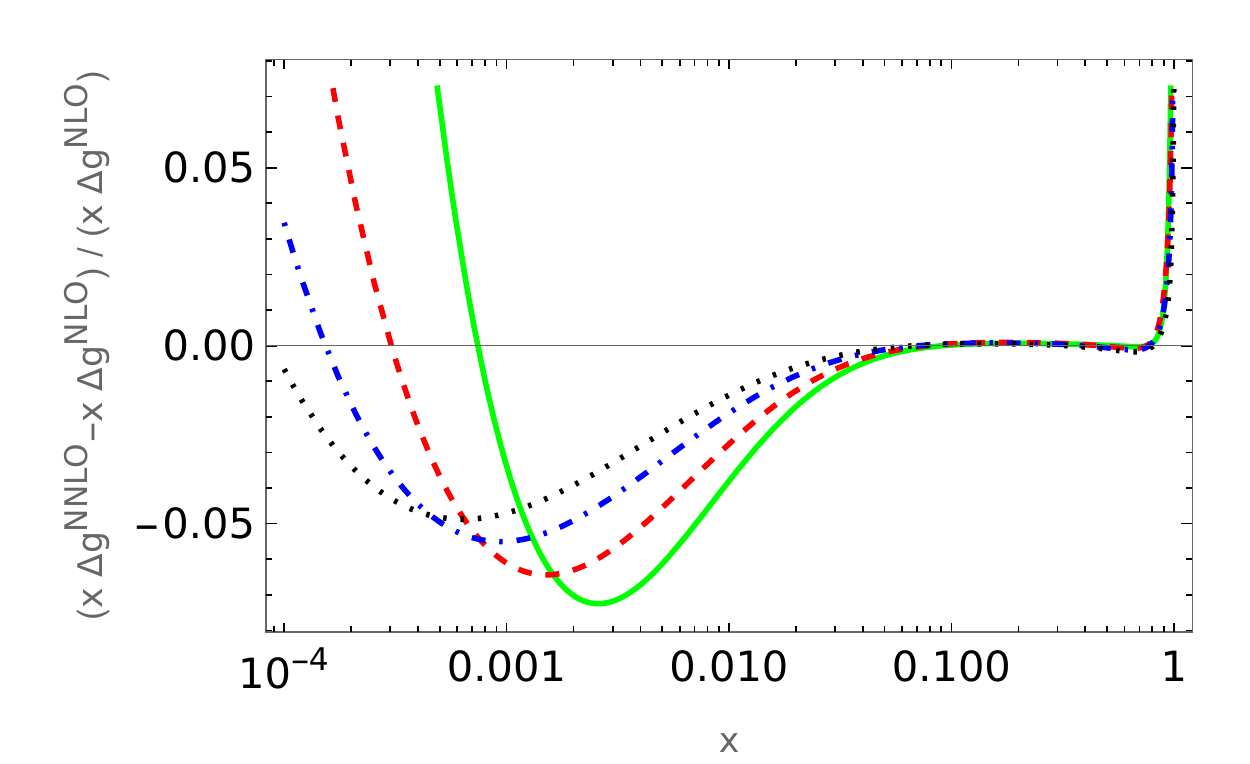} 
\caption{\sf  Ratios of the relative NNLO to NLO corrections of the 
singlet $(\Sigma)$ and gluon (G) distributions as functions of $x$. 
Upper panels: 
unpolarized case; lower panels: polarized case. 
$Q^2 = 10, 10^2 , 10^3 , 10^4~\GeV^2$: dotted, dash-dotted, dashed, 
full lines. From Ref.~\cite{Saragnese:2022uil}.}
\label{fig:1}
\end{figure}
One may now ask the question which order corrections are still important for current experimental 
precision analyses. Present day DIS data have an accuracy of up to O(1\%). Future data, e.g. at the 
EIC\cite{EIC}, will reach at least this level. As shown in Figure~\ref{fig:1} the NNLO corrections are 
not enough, in particular in the smaller $x$ and large $x$ regions. This applies also to the high 
luminosity data from the LHC. Therefore the four--loop splitting functions shall be calculated.
\section{Massless Wilson Coefficients}
\label{sec:5}

\vspace*{1mm}
\noindent
For the massless Wilson coefficients the one--loop corrections were given in
\cite{Furmanski:1981cw,Bodwin:1989nz}. The two--loop corrections were computed in
Refs.~\cite{WC2L,Larin:1991fv,Moch:1999eb} and the three--loop corrections were calculated
in Refs.~\cite{Vermaseren:2005qc,Moch:2008fj,Blumlein:2022gpp}. First color factor contributions
of $O(N_F^2)$ were computed recently in Ref.~\cite{Basdew-Sharma:2022vya} at four loops.
Up to the three--loop level all these quantities can be represented in terms of harmonic sums in Mellin 
space and the following 60
harmonic sums contribute\cite{Blumlein:2022gpp}, after algebraic reduction~\cite{Blumlein:2003gb}
\begin{eqnarray}
&&\{
S_1;
S_2,
S_{-2};
S_3,
S_{-3},
S_{2,1},
S_{-2,1};
S_4,
S_{-4},
S_{-2,2},
S_{3,1},
S_{-3,1},
S_{2,1,1},
S_{-2,1,1};
S_5,
\nonumber\\ &&
S_{-5},
S_{-2,3},
S_{2,3},
S_{2,-3},
S_{-2,-3},
S_{2,2,1},
S_{-2,1,-2},
S_{-2,2,1},
S_{4,1},
S_{-4,1},
S_{2,1,-2},
S_{3,1,1},
\nonumber\\ &&
S_{-3,1,1},
S_{2,1,1,1},
S_{-2,1,1,1};
S_6,
S_{-6},
S_{-3,3},
S_{4,2},
S_{4,-2},
S_{-4,2},
S_{-4,-2},
S_{5,1},
S_{-5,1},
\nonumber\\ &&
S_{-2,2,-2},
S_{-2,2,2},
S_{2,-3,1},
S_{-2,3,1},
S_{-3,1,-2},
S_{-3,-2,1},
S_{-3,2,1},
S_{-4,1,1},
S_{2,3,1},
S_{3,1,-2},
\nonumber\\ &&
S_{3,2,1},
S_{4,1,1},
S_{-2,-2,1,1},
S_{-2,1,1,2},
S_{-2,2,1,1},
S_{2,-2,1,1},
S_{2,2,1,1},
S_{3,1,1,1},
S_{-3,1,1,1},
\nonumber\\ &&
S_{2,1,1,1,1},
S_{-2,1,1,1,1}
\}.
\end{eqnarray}
The harmonic sums are recursively defined by
\begin{eqnarray}
S_{b,\vec{a}}(N) = \sum_{k = 1}^N \frac{({\rm sign}(b))^k}{k^{|b|}} S_{\vec{a}}(k),~~S_\emptyset = 1,
b, a_i \in \mathbb{Z} \backslash \{0\},~~N \in \mathbb{N} \backslash \{0\}. 
\end{eqnarray}
\section{Massive Wilson Coefficients}
\label{sec:6}

\vspace*{1mm}
\noindent
The massive Wilson coefficients receive single mass and two--mass contributions (due to both
charm and bottom quark corrections being present). We mainly will discuss asymptotic scales
$Q^2 \gg m_Q^2$ subsequently. In this case one obtains the following representation for the 
five contributing massive Wilson coefficients up to three--loop order\cite{Bierenbaum:2009mv}
\begin{eqnarray}
\label{eqWIL1}
L_{2,q}^{\sf NS}(N_F) &=&
a_s^2 \left[A_{qq,Q}^{{\sf NS}, (2)}(N_F) +
\hat{C}^{{\sf NS}, (2)}_{2,q}(N_F)\right]
\nonumber\\
&+&
a_s^3 \left[A_{qq,Q}^{{\sf NS}, (3)}(N_F)
+  A_{qq,Q}^{{\sf NS}, (2)}(N_F) C_{2,q}^{{\sf NS}, (1)}(N_F)
+ \hat{C}^{{\sf NS}, (3)}_{2,q}(N_F)\right]
\\
\label{eqWIL2}
{\tilde{L}}_{2,q}^{\sf PS}(N_F) &=&
a_s^3 \left[~\Atil_{qq,Q}^{{\sf PS}, (3)}(N_F)
+  A_{gq,Q}^{(2)}(N_F)~~ \Ctil_{2,g}^{(1)}(N_F+1)
+ \hat{\Ctil}^{{\sf PS}, (3)}_{2,q}(N_F)\right]
\\
\label{eqWIL3}
{\tilde{L}}_{2,g}^{\sf S}(N_F) &=&
a_s^2 A_{gg,Q}^{(1)}(N_F) \Ctil_{2,g}^{(1)}(N_F+1)
\nonumber\\ &+&
a_s^3 \Bigl[~\Atil_{qg,Q}^{(3)}(N_F)
+  A_{gg,Q}^{(1)}(N_F)~~\Ctil_{2,g}^{(2)}(N_F+1)
+  A_{gg,Q}^{(2)}(N_F)
\nonumber\\ && \hspace*{5mm}
~~\cdot \Ctil_{2,g}^{(1)}(N_F+1)
+  ~A_{Qg}^{(1)}(N_F)~~\Ctil_{2,q}^{{\sf PS}, (2)}(N_F+1)
+ \hat{\Ctil}^{(3)}_{2,g}(N_F)\Bigr]
\\
\label{eqWIL4}
H_{2,q}^{\sf PS}(N_F)
&=& a_s^2 \left[~A_{Qq}^{{\sf PS}, (2)}(N_F)
+~\Ctil_{2,q}^{{\sf PS}, (2)}(N_F+1)\right]
\nonumber\\
&+& a_s^3 \Bigl[~A_{Qq}^{{\sf PS}, (3)}(N_F)
+~\Ctil_{2,q}^{{\sf PS}, (3)}(N_F+1)
+ A_{gq,Q}^{(2)}(N_F)~\Ctil_{2,g}^{(1)}(N_F+1)
\nonumber\\ && \hspace*{5mm}
+ A_{Qq}^{{\sf PS},(2)}(N_F)~C_{2,q}^{{\sf NS}, (1)}(N_F+1)
\Bigr]
\\
\label{eqWIL5}
H_{2,g}^{\sf S}(N_F) &=& a_s \left[~A_{Qg}^{(1)}(N_F)
+~\Ctil^{(1)}_{2,g}(N_F+1) \right] \nonumber\\
&+& a_s^2 \Bigl[~A_{Qg}^{(2)}(N_F)
+~A_{Qg}^{(1)}(N_F)~C^{{\sf NS}, (1)}_{2,q}(N_F+1)
+~A_{gg,Q}^{(1)}(N_F)
\nonumber\\ &&
\hspace*{5mm}
~\cdot \Ctil^{(1)}_{2,g}(N_F+1)
+~\Ctil^{(2)}_{2,g}(N_F+1) \Bigr]
\nonumber\\ &+&
a_s^3 \Bigl[~A_{Qg}^{(3)}(N_F)
+~A_{Qg}^{(2)}(N_F)~C^{{\sf NS}, (1)}_{2,q}(N_F+1)
+~A_{gg,Q}^{(2)}(N_F)
\nonumber\\ &&
\hspace*{5mm}
~\cdot \Ctil^{(1)}_{2,g}(N_F+1)
+~A_{Qg}^{(1)}(N_F)\left[
C^{{\sf NS}, (2)}_{2,q}(N_F+1)
+~\Ctil^{{\sf PS}, (2)}_{2,q}(N_F+1)\right]
\nonumber\\ && \hspace*{5mm}
+~A_{gg,Q}^{(1)}(N_F)~\Ctil^{(2)}_{2,g}(N_F+1)
+~\Ctil^{(3)}_{2,g}(N_F+1) \Bigr],
\end{eqnarray}
where $\tilde{f}(N_F) = f(N_F)/N_F, \hat{f}(N_F) = f(N_F+1) - f(N_F)$. Here $C_i$ are the respective 
contributions of the massless Wilson coefficients and $A_{ij}^{(k)}$ are the massive $k$-loop order
OMEs. 
In the following we will deal with neutral--current interactions and the structure functions
$F_2(x,Q^2)$ and $F_L(x,Q^2)$ \cite{Blumlein:2006mh}. Also for charged current
processes higher order massive Wilson coefficients have been calculated.
It turns out that beyond two--loop order several new mathematical quantities beyond the harmonic 
sums are contributing, cf.~Section~\ref{sec:9}.
\subsection{Single mass corrections}
\label{sec:61}

\vspace*{1mm}
\noindent
The one--loop corrections can be calculated for general values of $Q^2$ and were obtained in 
Refs.~\cite{HQone} in the unpolarized and polarized cases. The tagged--heavy flavor two--loop 
corrections
were calculated numerically in Refs.~\cite{Tag2}. Note that these corrections do not refer
to the inclusive structure functions.  These were calculated in the case of asymptotic scales
$Q^2 \gg m_Q^2$ in Refs.~\cite{Buza:1995ie,Buza:1996xr,Bierenbaum:2007qe,Bierenbaum:2008yu,
Blumlein:2016xcy,Bierenbaum:2022biv}. 
The flavor non--singlet contributions can also be obtained in closed form for general values of 
$Q^2$, cf.~\cite{Buza:1995ie,Buza:1996xr,Blumlein:2016xcy}. This is also the case for 
the pure singlet contributions \cite{TLPS} and one may obtain a systematic expansions of the 
contributing power corrections of $O((m_Q^2/Q^2)^k)$. Here root--valued alphabets play a role 
and the results are given by incomplete elliptic integrals in part which are iterative 
integrals, unlike complete elliptic integrals. The analytic asymptotic two--loop results depend 
only on 
harmonic sums \cite{HSUM}, as do the logarithmic scale corrections to three--loop order. 
The latter corrections were obtained in Refs.~\cite{LOG3}. The three--loop corrections
to the unpolarized asymptotic Wilson coefficients were computed in Refs.~\cite{WC3}
and in the polarized case in Refs.~\cite{LOG3,Ablinger:2019etw}

Massive OMEs determine also the transition matrix elements in the variable flavor scheme 
(VFNS). The corrections up to two--loop order were calculated in 
Refs.~\cite{Buza:1996wv,Bierenbaum:2009zt,Bierenbaum:2022biv}  and the single and 
two--mass VFNS were given in  \cite{Buza:1996wv,Blumlein:2018jfm,Bierenbaum:2022biv}. At 
three--loop order the massive OMEs beyond those contributing to the massive Wilson 
coefficients, were calculated in Refs.~\cite{LOG3,OMEbU,Ablinger:2014lka} for the 
unpolarized case and in 
Refs.~\cite{LOG3,Behring:2021asx,OMEbU} in the polarized case.

The transition relations in the single mass variable flavor number scheme\cite{Buza:1996wv} are given by
\begin{eqnarray}
\label{HPDF1}
\lefteqn{
f_k(N_F+1,\mu^2,m^2,N) + f_{\overline{k}}(N_F+1,\mu^2,m^2,N) =} \nonumber\\
&& A_{qq,Q}^{\rm NS}\left(N_F,\frac{\mu^2}{m^2},N\right)
\cdot \bigl[f_k(N_F,\mu^2,N)
+ f_{\overline{k}}(N_F,\mu^2,N)\bigr]
 + \tilde{A}_{qq,Q}^{\rm
PS}\left(N_F,\frac{\mu^2}{m^2},N\right)
\nonumber\\ &&
\cdot \Sigma(N_F,\mu^2,N)
 + \tilde{A}_{qg,Q}\left(N_F,\frac{\mu^2}{m^2},N\right)
\cdot G(N_F,\mu^2,N),
\\
\label{fQQB}
\lefteqn{f_Q(N_F+1,\mu^2,m^2,N) + f_{\overline{Q}}(N_F+1,\mu^2,m^2,N) =} \nonumber\\
&&
{A}_{Qq}^{\rm PS}\left(N_F,\frac{\mu^2}{m^2},N\right)
\cdot \Sigma(N_F,\mu^2,N)
+ {A}_{Qg}\left(N_F,\frac{\mu^2}{m^2},N\right)
\cdot G(N_F,\mu^2,N)
\end{eqnarray}
\begin{eqnarray}
\hspace*{-2cm}
\Sigma(N_F+1,\mu^2,m^2,N) &=& \Biggl[A_{qq,Q}^{\rm NS}\left(N_F, \frac{\mu^2}{m^2},N\right) +
          N_F \tilde{A}_{qq,Q}^{\rm PS}\left(N_F, \frac{\mu^2}{m^2},N\right)
\nonumber\\ &&
         + {A}_{Qq}^{\rm PS}\left(N_F, \frac{\mu^2}{m^2},N\right)
        \Biggr]
\cdot \Sigma(N_F,\mu^2,N) \nonumber\\ &&
+ \Biggl[N_F \tilde{A}_{qg,Q}\left(N_F,
\frac{\mu^2}{m^2},N\right) 
+
          {A}_{Qg}\left(N_F, \frac{\mu^2}{m^2},N\right)
\Biggr]
\nonumber\\
&&
\cdot G(N_F,\mu^2,N)
\\
\Delta(N_F+1,\mu^2,m^2,N) &=&
  f_k(N_F+1,\mu^2,N)
+ f_{\overline{k}}(N_F+1,\mu^2,m^2,N)
\nonumber\\ &&
- \frac{1}{N_F+1}
\Sigma(N_F+1,\mu^2,m^2,N)\\
\label{HPDF2}
G(N_F+1,\mu^2,m^2,N) &=& A_{gq,Q}\left(N_F,
\frac{\mu^2}{m^2},N\right)
                    \cdot \Sigma(N_F,\mu^2,N)
\nonumber\\
&&
+ A_{gg,Q}\left(N_F, \frac{\mu^2}{m^2},N\right)
                    \cdot G(N_F,\mu^2,N)~.
\end{eqnarray}
In this way one may define also heavy quark parton densities $f_{Q(\overline{Q})}$ in the region 
$\mu^2 \gg m_Q^2$. In the two--mass case one may separate the genuine two--mass contributions
to $f_c$ and $f_b$, cf.~\cite{Ablinger:2017err}.
\subsection{Two-mass 
corrections} \label{sec:62} 

\vspace*{1mm}
\noindent
The two--mass corrections to the different massive OMEs, except that of $(\Delta) 
A_{Qg}^{(3)}$, can be calculated in terms of iterative integrals using square--root valued 
alphabets in which the real mass--ratio  $\eta = m_c^2/m_b^2$ appears in $x$--space. In 
addition new special constants are contributing associated to these functions. The corrections, 
except those for $A_{Qg}^{(3)}$ in the unpolarized case, were calculated in 
Refs.~\cite{Ablinger:2017err,Ablinger:2018brx,Ablinger:2017xml}. In the polarized case the 
three--loop corrections were computed in Refs.~\cite{Ablinger:2017err,Ablinger:2020snj,
Ablinger:2019gpu}. The VFNS in the two--mass case has been given in \cite{Ablinger:2017err}. It extends 
the one given in Section~\ref{sec:61} and accounts for the fact that the mass ratio $m_c^2/m_b^2$ is not 
small.
\section{Scheme-invariant evolution}
\label{sec:7}

\vspace*{1mm}
\noindent
The systematically and theoretically best way to measure the strong coupling constant
in deep--inelastic scattering is due to the evolution of a given structure function itself.
This requires specific experimental conditions, which were sometimes not available 
at some of the deep--inelastic facilities in the past.
Having proton and deuteron data available in the same $(x, Q^2)$--bins and performing
the deuteron wave-function corrections allows to measure the following non--singlet 
structure  functions
\begin{eqnarray}
F_2^{\rm NS}(x,Q^2) &=& F_2^p - F_2^d =  \frac{x}{6} C_q^{\rm NS +} \otimes
[u + \bar{u} - d - \bar{d}],
\\
x g_1^{\rm NS}(x,Q^2) &=& x(g_1^p - g_1^d) = \frac{x}{6} \Delta C_q^{\rm NS +} \otimes
[\Delta u + \Delta \bar{u} - \Delta d - \Delta \bar{d}],
\end{eqnarray}
see Ref.~\cite{Blumlein:2021lmf}\footnote{Scheme invariant evolution equations in the singlet case were 
considered in Refs.~\cite{Furmanski:1981cw,SINVS}}. The massless and the massive non--singlet Wilson 
coefficients are
available to three--loop order \cite{Vermaseren:2005qc,Blumlein:2022gpp,Ablinger:2014vwa},
including the two--mass corrections \cite{Ablinger:2017err}. The scale evolution  of the non--singlet 
combination of the parton distribution functions forming a {\it single} input density, requires 
four--loop anomalous dimensions. The investigation of moments shows,
that these quantities can be extremely well constrained by a Pad\'e--approximant of the lower order 
anomalous dimensions, implying a negligible theory error. The above equations can be rewritten in
terms of evolution operators $(\Delta) E^{\rm NS}$ 
\begin{eqnarray}
\label{eq:SI1}
F_2^{\rm NS}(x,Q^2) &=& E^{\rm NS}(x, Q^2, Q_0^2) \otimes F_2^{\rm NS}(x,Q^2_0), 
\\
\label{eq:SI2}
xg_1^{\rm NS}(x,Q^2) &=& x[\Delta E^{\rm NS}(x, Q^2, Q_0^2) \otimes g_1^{\rm NS}(x,Q^2_0)]. 
\end{eqnarray}
Here the evolution operators can be analytically calculated in Mellin space in the analyticity 
region 
of $N \in \mathbb{C}$. The $x$--space result is then obtained by a single numerical contour integral
around the singularities of the problem, cf.~\cite{Diemoz:1987xu,Blumlein:1997em}.
Measuring the input structure functions at $Q_0^2$ with correlated errors, the evolution 
from $Q_0^2$ to the higher scales $Q^2$ depends only on a single parameter, the strong coupling 
constant $a_s(M_Z^2)$ or the QCD--scale $\Lambda_{\rm QCD}$. The charm quark mass may be fixed within 
errors in this process and accounted for by error 
propagation. A measurement of this kind is proposed for further facilities, like the EIC \cite{EIC} 
or the LHeC \cite{LHEC}.
\section{The Drell--Yan process}
\label{sec:8}

\vspace*{1mm}
\noindent
The Drell--Yan process of hadronic lepton pair production $pp \rightarrow \gamma^*/Z^* + X$
with subsequent leptonic decay of the virtual gauge bosons \cite{Drell:1970wh} or the associated charged 
current processes probe quark--antiquark initial states at leading order. Therefore this process is
particularly sensitive to the sea quark distributions and yields complementary information to 
deep--inelastic scattering in disentangling the different light flavor distributions.
The one--loop corrections to this process were calculated in Refs.~\cite{DY1} around 1980.
The two--loop corrections have been completed 1990 in \cite{Hamberg:1990np}. A subset of the Wilson 
coefficients is also related to the initial state QED corrections of $e^+ e^- \rightarrow \gamma^*/Z^* + 
X$, for massive electrons in the limit $m_e^2/s \rightarrow 0$, where $s$ denotes the cms energy, cf.
Ref.~\cite{Blumlein:2020jrf}. Like for all massless and massive two--loop single scale Wilson 
coefficients it has been shown in Ref.~\cite{Blumlein:2005im} that also in the case of the unpolarized 
and 
polarized Drell--Yan processes and  Higgs--boson production only six functions are needed in Mellin $N$ 
space to describe these quantities. Here only harmonic sums \cite{HSUM} contribute. The three--loop 
corrections were  calculated in Refs.~\cite{Duhr:2020seh}. Here also elliptic integrals contribute to the 
scattering cross section, if expressed in the variable $\hat{s}/s$, where $\hat{s}$ is the cms energy
of the virtual gauge boson.
In the experimental analysis one has to use differential distributions, such as encoded in the 
packages {\tt DYNNLO, FEWZ, MATRIX, MCFM} \cite{DIFFDY}.
\section{Conclusions}
\label{sec:9}

\vspace*{1mm}
\noindent
Perturbative QCD has evolved significantly over the last 50 years and proven to be the correct
theory of the strong interactions at high virtualities. While reviews like Ref.~\cite{Abers:1973qs} in 
1973 still were 
reluctant to evoke $SU_{3c}$ as part of the Standard Model, QCD allows now for highly precise 
predictions.

These analytic results required new mathematical and computer--algebraic technologies to be obtained.
On the side of computer algebra we would like to mention in particular the IBP methods 
\cite{IBP}, {\tt Forcer} \cite{Ruijl:2017cxj}, the packages {\tt Sigma} \cite{SIG} and 
{\tt HarmonicSums} \cite{HARMSU}, the method of arbitrary high moments 
\cite{Blumlein:2017dxp}, and the method to perform the inverse Mellin transform without 
giving an explicit general $N$ expression \cite{Behring:2023rlq}. At 
the mathematics side new developments have set in around 1998 
with harmonic sums \cite{HSUM}, generalized harmonic sums \cite{GHSUM}, cyclotomic harmonic 
sums \cite{Ablinger:2011te}, finite and infinite binomial sums 
\cite{Ablinger:2014bra,INVBIN}, related iterated integrals 
\cite{Remiddi:1999ew,GHSUM,Ablinger:2011te,Ablinger:2014bra},
special numbers, e.g.~Ref.~\cite{Blumlein:2009cf}, and methods related to $_2F_1$--solutions 
\cite{Ablinger:2017bjx} and complete elliptic 
integrals \cite{Ablinger:2017bjx,ELLIPT}. For a 
survey on these methods see Refs.~\cite{Blumlein:2018cms}.
Here the main question is: What can be integrated 
analytically and how? An important aspect in this context is anti-differentiation 
\cite{Blumlein:2021ynm}.
This development is still ongoing and we look forward for the new brilliant results to come.

Finally, we would like to comment on fundamental parameters of the Standard Model, such as
$\alpha_s(M_Z^2)$ and $m_c$, which can already be determined by the present high--precision data.
These still will be improved when the calculation for all the three--loop heavy flavor 
corrections are completed.
In earlier analyses we obtained in the non--singlet and singlet cases the following values
for $\alpha_s(M_Z^2)$\footnote{Working under comparable cuts as ours, also in 
Ref.~\cite{Thorne:2013hpa,Dulat:2015mca}  lower values of $\alpha_s^{\rm N^2LO}(M_Z^2) = 
0.1136$ and $0.1150 {\tiny \begin{array}{c} + 0.0060 \\ - 0.0040 \end{array}}$
were obtained. Furthermore, we agree with the results of Ref.~\cite{Jimenez-Delgado:2014twa}, 
$\alpha_s^{\rm N^2LO}(M_Z^2) = 0.1136 \pm 0.0004$. Also a series of other measurements 
at N$^2$LO deliver a series of values significantly below the world average, cf. 
Ref.~\cite{alphas}.}
\begin{alignat}{4}
\alpha_s^{\rm N^3LO, NS}(M_Z^2) &~=~& 0.1141 \pm 0.0022  &~~[13]&,
\\
\alpha_s^{\rm N^2LO}(M_Z^2) &~=~& 0.1140 \pm 0.0009      &~~[12]&,
\end{alignat}
and the charm quark mass
\begin{eqnarray}
m_c(m_c) &=& 1.252 \pm 0.018~\GeV.~~~[12]. 
\end{eqnarray}
Note that there is still a $0.07~\GeV$ theory error involved in the latter, which will
be significantly reduced after the massive three--loop corrections are completely available. The result 
is very well compatible with the four--loop result based on $e^+ e^-$ annihilation data
\begin{eqnarray}
m_c(m_c) &=& 1.279 \pm 0.008~\GeV.~~~[130]. 
\end{eqnarray}
Dedicated measurements at future high luminosity DIS facilities, like the EIC\cite{EIC} and 
LHeC\cite{LHEC}, are believed to improve these results further and to finally resolve the problem
of still conflicting results on $\alpha_s(M_Z^2)$ from different measurements 
\cite{alphas,BETHKE}.


\end{document}